\documentclass{ELSP}
\PassOptionsToPackage{unicode}{hyperref}
\PassOptionsToPackage{hyphens}{url}
\IfFileExists{upquote.sty}{\usepackage{upquote}}{}
\IfFileExists{microtype.sty}{% use microtype if available
	\usepackage[]{microtype}
	\UseMicrotypeSet[protrusion]{basicmath} % disable protrusion for tt fonts
}{}
\makeatletter
\IfFileExists{parskip.sty}{%
	\usepackage{parskip}
}
\makeatother
\IfFileExists{xurl.sty}{\usepackage{xurl}}{} % add URL line breaks if available
\IfFileExists{bookmark.sty}{\usepackage{bookmark}}{\usepackage{hyperref}}
\hypersetup{
	pdfcreator={ELSP}}
\usepackage{longtable,booktabs,array}
\usepackage{calc} % for calculating minipage widths
\usepackage{etoolbox}
\makeatletter
\patchcmd\longtable{\par}{\if@noskipsec\mbox{}\fi\par}{}{}
\makeatother
\IfFileExists{footnotehyper.sty}{\usepackage{footnotehyper}}{\usepackage{footnote}}
\makesavenoteenv{longtable}
\usepackage{setspace}
\usepackage{graphicx}
\usepackage{amsmath}
\usepackage{times}
\usepackage{mathptmx}
\DeclareMathAlphabet{\mathbfit}{OML}{txmi}{b}{it}
\usepackage{color}
\usepackage{caption}
\usepackage{subcaption}
\usepackage{geometry}
\usepackage{float}
\usepackage{enumerate}
\usepackage{enumitem}
\usepackage{stfloats}
\usepackage{ragged2e}
\usepackage{titlesec}
\usepackage{ragged2e}
\usepackage{lscape}
\usepackage{multirow}

\sloppypar

\titleformat{\section}[hang]
  {\fontsize{12pt}{12pt}\selectfont\bfseries\color[RGB]{0,131,255}} 
  {\thesection}{0.5em}{}

\titleformat{\subsection}[hang]
  {\fontsize{12pt}{12pt}\selectfont\itshape} 
  {\thesubsection}{0.5em}{}

\titleformat{\subsubsection}[hang]
  {\fontsize{12pt}{12pt}\selectfont} 
  {\thesubsubsection}{0.5em}{}

\usepackage{fancyhdr}
\fancypagestyle{firstpage}
{
	\fancyhf{}
	\setlength{\headsep}{6px}
}

\hypersetup{
    colorlinks=false, % 
    pdfborder={0 0 0}, % 
    hidelinks % 
}
\usepackage[utf8]{inputenc}
\usepackage[table,xcdraw,RGB]{xcolor}
\begin{document}
%\hfill \textbf{\large Type of the paper}
\thispagestyle{firstpage}

\setstretch{1.24}
\begin{flushleft}
% \papertype{paper type}
%{\sffamily \small \noindent {\textls[-12]{Article $\mid$ Received 11 June 2026; Revised 20 July 2026; Accepted 23 July 2026; Published XX August 2026}}}\\[-0.7ex]
%Example
%\timeline{15 May 2022}{17 July 2022}{20 September 2022}
%{\sffamily\small{https://doi.org/10.55092/asymmetry20260005}}

\vspace{8pt}
{\raggedright
\papertitle{%
  Matching of perturbative and exponentiated initial state
radiation corrections to $e^+e^-$-annihilation
}}

\vspace{12pt}
\hspace{-0.1cm}\authorname{Andrej} {Arbuzov} {1,2,*},
\hspace{-0.1cm}\authorname{Uliana} {Voznaya} {1,2}

\vspace{12pt}
\formatintroduction{1}{Bogoliubov Laboratory for Theoretical Physics, Joint Institute for Nuclear Research, Dubna, Russia}
\formatintroduction{2}{Institute of Engineering and Physics, Dubna State University, Dubna, Russia}

\vspace{12pt}
\authoremail{Correspondence author} {arbuzov@theor.jinr.ru.}
\end{flushleft}

\vspace{10pt}
\noindent\textbf{\textbf{\textcolor[RGB]{0,131,255}{Abstract:}}} The behavior of higher-order radiative corrections due to initial state radiation in processes
\textls[+12]{of electron-positron annihilation is analyzed. Numerical results for energies of future colliders are presented.} Uncertainties of the known results on these corrections are estimated. A modified scheme for simultaneous exponentiation of pure photonic and non-singlet pair corrections is presented. 
Matching of the exponentiated results with the existing analytic higher-order calculations is constructed. 
A new subtraction scheme is suggested. The results obtained are important for increasing the relative accuracy of the theoretical description of processes that will be studied in future electron-positron colliders to a level of approximately $10^{-5}$.

\vspace{+12pt}
\noindent\textbf{\textcolor[RGB]{0,131,255}{Keywords:}}  radiative corrections; exponentiation; parton distribution functions; QED; electron-positron annihilation

\section{Introduction}

Precise theoretical predictions of radiative corrections to electron-positron annihilation are important for experiments at future electron-positron colliders like FCC-ee~\cite{FCC:2018evy} and CEPC~\cite{CEPCStudyGroup:2018ghi}. 
Despite the great progress in the development of multi-loop calculation methods, computing higher-order radiative corrections to realistic differential observables in high-energy physics remains a complex task. Meanwhile, the accuracy of the analysis of many observables, first of all related to precision tests of the Standard Model and extraction of its parameters, suffers from uncertainties
in the theoretical description of effects due \textls[-12]{to initial state radiation~\cite{FCC:2018byv}. One of the methods that allows evaluation of higher-order corrections without} their direct calculations is exponentiation of known lowest-order results, which performes a reorganization of the perturbative expansion and resummation of certain contributions to all orders. Another powerful method is the so-called QED structure function
% we commented about the abbreviation "QED" above
 approach~\cite{Kuraev:1985hb}, which allows one to calculate systematically the part of higher-order corrections enhanced by powers of large logarithms. In this article, we discuss matching of these approaches and estimate the corresponding theoretical uncertainties.

\textls[+18]{Exponentiation in particle physics has been studied since 1960s. In the work of D.R.~Yennie,} S.C.~Frautschi, and H.~Suura~\cite{Yennie:1961ad} infrared divergences were resummed into a universal exponential factor. In the Yennie-Frautschi-Suura (YFS) exponentiation formalism, all photons emitted in processes like electron scattering by a potential, bremsstrahlung off nucleons, \textit{etc.}, are separated into soft and hard parts based on their energy, and radiation of soft and virtual photons is associated with infrared singularities, which cancel each other. The remaining contribution can be re-summed in an exponential form. There are also some Monte-Carlo (MC) simulation programs based on the YFS method: PHOTOS~\cite{Golonka:2005pn}, \mbox{KKMC \cite{JADACH2000260}}, \textls[-12]{BHLUMI~\cite{Jadach:1991by, Placzek:2026hvx}}, Sherpa~\cite{Krauss:2022ajk}, and others.
Exponentiation of the pure photonic part of the electron structure function (parton distribution function) was derived by V.N.~Gribov and L.N.~Lipatov~\cite{Gribov:1972rt}. E.A.~Kuraev and V.S.~Fadin~\cite{Kuraev:1985hb} suggested an ad
hoc exponentiation procedure which allows to improve accuracy of the finite order perturbative solution
and to take into account a part of pair corrections. In the work by G.~Passarino~\cite{Passarino:2001wy}, connection between exponentiation and structure functions is discussed in detail. \textls[-18]{Amplitude-based resummation in collinear and infrared (IR)
 limits and its YFS MC realization was discussed in the work~\cite{Ward:2026bco}, 
and Coherent Exclusive Exponentiation (CEEX)~\cite{Jadach:1998jb} and Exclusive Exponentiation (EEX)~\cite{Jadach:1989nm}} realization was discussed in~\cite{Frixione:2022ofv}.

\textls[-46]{The method of structure functions was introduced first for QED in the work of V.N.~Gribov and L.N.~\mbox{Lipatov \cite{Gribov:1972ri}}, where the electron structure functions were obtained. 
The Dokshitzer--Gribov--Lipatov--Altarelli--Parisi} 
evolution equations \cite{Altarelli:1977zs,Dokshitzer:1977sg} were derived for QCD
% QCD is a commonly accepted abbreviaton, we think the full form is not necessary
 based on the latter approach, and they were reduced back to the QED case by E.A.~Kuraev and V.S.~Fadin~\cite{Kuraev:1985hb}. 
There are numerous applications and further developments of the method
within the leading logarithmic approximation, see e.g.
references~\cite{Nicrosini:1986sm,Skrzypek:1992vk,Cacciari:1992pz,WorkingGrouponRadiativeCorrections:2010bjp}. 
Application of the method in the next-to-leading order (NLO) approximation was demonstrated for the first time in~\cite{Berends:1987ab} 
for derivation of QED radiative corrections due to initial state radiation (ISR) in electron-positron annihilation. 
Then it was applied for calculations of $\mathcal{O}(\alpha^2L)$ corrections
to a few other processes including muon decay, deep inelastic scattering~\cite{Blumlein:2002fy}, Bhabha scattering, and charge and helicity asymmetries.
The structure function approach allows avoiding direct loop calculations. But only the most numerically \textls[-18]{significant corrections are taken into account. These are corrections enhanced by the so-called large logarithm}
\begin{equation}
L = \ln \frac{\mu_F^2}{\mu_R^2},
\end{equation}
where $\mu_F$ is factorization scale and $\mu_R$ is renormalization scale. 
The natural choice of $\mu_R$ in QED is the electron mass, and $\mu_F$ should correspond to a
characteristic (high) energy of the process. 

\textls[-18]{An alternative approach is the soft-collinear effective theory (SCET) which allows to separate soft and hard collinear contributions in an easier way~\cite{Tomalak:2021hec, Tomalak:2022xup, Hill:2016gdf}. Nevertheless, here we use the structure (parton density) function method which systematically provides higher-order contributions enhanced by the large logarithms.}

The article is organized as follows. In the next Section, ISR radiative corrections to electron-positron annihilation are discussed. In the third Section, we describe exponentiation procedures and suggest an expression for exponentiation of parton distribution function corresponding to both photonic and non-singlet pair corrections. The fourth Section is dedicated to the scheme dependence of the corrections. In the fifth Section, theoretical uncertainties are estimated and discussed.

\section{ISR radiative corrections}

Here we apply the parton distribution function (structure function) approach~\cite{Kuraev:1985hb}. 
The cross section of electron-positron annihilation into a virtual photon or $Z$ boson, decaying then, say, into muon pair
\vspace{-4pt}
\begin{equation}
e^+e^- \to \gamma^{*}(Z^{*}) \to \mu^+\mu^-
\end{equation}

\vspace{-8pt}
\noindent with ISR corrections can be represented in the form of the convolution of two electron parton distribution functions (PDFs) and the partonic cross section~\cite{Berends:1987ab}
\vspace{-4pt}
\begin{eqnarray} \label{master}
&&	\frac{d\sigma^{\mathrm{NNLL}}_{\bar{e} e}(s')}{ds'} = \sum \limits_{i,j= e\!, \bar{e}\!,  \gamma}  \int \limits^{1}_{\bar{z}_1} \int \limits^{1}_{\bar{z}_2} d z_1 
	d z_2 D_{i e} \left(z_1,\frac{\mu_R^2}{\mu^2_F}\right)  D_{j \bar{e}} \! \left( \! z_2,\frac{\mu_R^2}{\mu^2_F} \! \right) \!
	\nonumber \\
&&	\times
	\left( \sigma^{(0)}_{ij} (s z_1 z_2) + \overline{\sigma}^{(1)}_{ij} (s z_1 z_2) +
    \overline{\sigma}^{(2)}_{ij} (s z_1 z_2) 
%	+ \mathcal{O}(\alpha^2 L^0)
    \right)
    \delta(s' \! - \! sz_1z_2)
	+ \mathcal{O}\left(\frac{m_e^2}{s}\right),
\end{eqnarray}

\vspace{-4pt}
\noindent where $\bar{e}\equiv e^+$ is positron and $e\equiv e^-$ is electron, $D_{i e}$ and $D_{j \bar{e}}$ are PDFs which define the probability density to find massless partons inside the massive electron (positron), $\sigma^{(0,1,2)}_{ij}$ are the Born~$(0)$, one-loop~$(1)$, and two-loop $(2)$ contributions to cross-sections of annihilation into $\gamma^*(Z^*)$ at the partonic level, $s$ is the initial center-of-mass energy squared, $s'$ is the invariant mass of the produced virtual photon (or $Z$-boson) squared, and $z$ is the energy fraction: $s'=sz$, $z\equiv z_1z_2$. Bars over $\overline{\sigma}^{(1,2)}_{ij}$ mean that these contributions are computed for massless partons within the $\overline{\mathrm{MS}}$ scheme for subtraction of electron mass singularities.
For consistency, the PDFs are also defined in the $\overline{\mathrm{MS}}$ scheme.

The expansion of the master equation for cross-section takes into account
the QED radiative corrections enhanced by the large logarithms and reads
\begin{eqnarray} \label{cij}
&&	\frac{d\sigma^{\mathrm{NNLL}}_{\bar{e} e}(s')}{ds'}  =  \frac{\sigma^{(0)}_{\bar{e} e}(s')}{s} \biggl\{ \delta(1-z) + \sum \limits^{\infty}_{\substack{k=1 \\ k\geq l \geq k-2}} 
	c_{kl}(z) \left(\frac{\alpha}{2\pi}\right)^k  L^{l}       
     + \mathcal{O}(\alpha^kL^{k-3}) \biggr\},
\end{eqnarray}
\textls[-8]{where $c_{kl}(z)$ are the coefficients to be computed. Here the sum can be divided into tree parts that consist of the leading (LL)}, next-to-leading (NLL), and next-to-next-to-leading (NNLL) logarithmic contributions \textls[+18]{($l=k$, $l=k-1$, and $l=k-2$, respectively). Higher-order coefficients $c_{kl}(z)$ in the leading and next-to-leading logarithmic approximations up to $c_{65}$ were first calculated in~\cite{Ablinger:2020qvo}},
and most of those coefficients were independently recalculated and corrected in~\cite{Arbuzov:2024tac}.
Corrections to the results for $c_{20}$ given in~\cite{Berends:1987ab} can be found in~\cite{Blumlein:2019srk} and~\cite{Hamberg:1990np}. 

\textls[+18]{To get the total cross section we have to integrate the differential cross section over $z$ from $z_{min}$ corresponding to a certain minimal value of the invariant mass of the muon pair defined by experimental conditions,}
\vspace{-8pt}
\begin{equation}
\sigma_{e^+ e^-} = \sigma^{(0)} (s) + \sum\limits_{\substack{k=1 \\ l \leq k}} \left(\frac{\alpha}{2\pi}\right)^k L^l \left[ \ \int \limits_{z_{min}}^{1 - \Delta} dz  \sigma^{(0)}(sz) c^{\Theta}_{kl} (z) + c^{\Delta}_{kl}\sigma^{(0)}(s)  \right].
\end{equation}

\vspace{-6pt}
\noindent \textls[+16]{Here $c^{\Theta}_{kl}$ and $c^{\Delta}_{kl}$ are the so-called $\Theta$ and $\Delta$ parts of the coefficients $c_{kl}$, see details in~\cite{Arbuzov:2024tac}. The first} corresponds to contributions from hard emission, and the second one corresponds to soft and virtual corrections. $\Delta$ is a small parameter dividing hard and soft contributions and is taken equal to $10^{-7}$ or $10^{-8}$. The final result doesn't depend on $\Delta$ for $\Delta\to 0$.
We present numerical results for relative corrections $h_{kl}$ in percent separately in different orders in $\alpha$ and $L$:
\vspace{-4pt}
\begin{equation}
h_{kl}   
  = \left(\frac{\alpha}{2\pi}\right)^k L^l \left( \int \limits_{z_{min}}^{1 - \Delta} dz  \sigma^{(0)}(sz) c^{\Theta}_{kl} (z) + c^{\Delta}_{kl} \sigma^{(0)}(s) \right) / \sigma^{(0)} (s) \cdot 100 \%.
\end{equation}

\vspace{-4pt}
We estimated numerical values of the leading and next-to-leading logarithmic corrections to initial state radiation in $e^+e^-$-annihilation at different energies $\bigl(\frac{M_Z - 1}{2}$, $\frac{M_Z}{2}$, $\frac{M_Z + 1}{2}$, $160$ GeV, $240$ GeV and $3$~TeV$\bigr)$ and different values of $z_{min}$ (see Table~\ref{tab:app1}). %Tables 2–4 are renamed Table A1, since they are extended data for one table caption and appear in the Appendix. Please check. -- It is fine.
At $160$~GeV and $240$~GeV a radiative return to $Z$-resonance can occur due to the effective reduction of the collision energy because of radiation from the initial state, which makes the values of the corrections at $z_{min}=0.1$ up to ten times larger than the same corrections at $\sqrt{s} = M_Z$. At $\sqrt{s}=240$~GeV the return to the resonance is at $z \approx 0.14$, where $sz=M_Z^2$, and at $\sqrt{s}=160$~GeV it is at $z \approx 0.32$. 

We work within the collinear factorization framework, so we integrate over the angles of emitted photons. Note that the angular distribution of bremsstrahlung photons is peaked in the direction collinear to their parent charged particle momenta. In average, the angle of the emitted photons doesn't affect sufficiently inclusive observables such as the total cross section and the differential distribution in the invariant mass of the final-state fermion pair. For exclusive observables, one has to run a full Monte Carlo simulation and the collinear approximation can be used only as an estimate (adequately accurate Monte Carlo codes are not yet available).

\section{Exponentiation}

Exponentiation procedure allows resummation of a part of corrections that are most important in
the domain of collinear and soft (or virtual) radiation. Exponentiation and resummation of QED corrections is discussed in details in~\cite{Passarino:2001wy}.

Exponentiated pure photonic part of the electron PDF $D_{ee}$ is given by the exact solution of the QED Dokshitzer-Gribov-Lipatov-Altarelli-Parisi (DGLAP)
 evolution equation found by V.N.~Gribov and L.N.~Lipatov in~\cite{Gribov:1972rt}:
\vspace{-4pt}
\begin{equation} \label{GL}
D_{ee}^{(exp, \gamma)}(z;\beta) = \frac{\beta}{2} \frac{(1-z)^{\frac{\beta}{2}-1}}{\Gamma(\frac{\beta}{2}+1)} \exp \biggl(\frac{\beta}{2} \left( \frac{3}{4} - \gamma_E \right)\biggr),
\end{equation}

\vspace{-4pt}
\noindent for $\Theta-$part, where $\beta =  \frac{2 \alpha}{\pi} (L -1)$, $\Gamma$ is the Euler Gamma-function, $\gamma_E \approx 0.577$ is the Euler-Mascheroni constant. We also need the $\Delta-$part for numerical applications:
\vspace{-4pt}
\begin{equation} \label{dexp}
D_{ee}^{(exp, \gamma) \Delta} \equiv \int\limits_{1-\Delta}^{1}d z D_{ee}^{(exp,\gamma)}(z;\beta)= \exp \left( \frac{\beta}{2} \ln \Delta  +\frac{3 \beta}{8}\right) \frac{\exp (-\gamma_E \beta /2)}{\Gamma(1+\beta/2)}.
\end{equation}

\vspace{-4pt}
As soon as the cross section has the structure $\sigma \otimes D \otimes D$, see Equation~(\ref{master}), we need to calculate the convolution of $D_{ee}^{(exp,\gamma) \otimes 2} = D_{ee}^{(exp,\gamma)} \otimes D_{ee}^{(exp,\gamma)}$ analytically. Technically it is very complicated, but we can present the $D_{ee}^{(exp,\gamma) \otimes 2}$ as
\vspace{-4pt}
\begin{equation}
D_{ee}^{(exp,\gamma) \otimes 2} (z; \beta)\approx D_{ee}^{(exp,\gamma)} (z; 2 \beta)\label{D2beta}.
\end{equation}

\vspace{-10pt}
\noindent This approximation becomes exact at $z \rightarrow 1$.

We suggest matching of the known perturbative higher-order corrections with the exponentiated result in the following form of an improved cross-section: 
\vspace{-4pt}
\begin{equation} 
\sigma^{imp}(z) = \sigma^{pert}(z) - D^{(exp,\gamma) \otimes 2}_{ser}(z) \ \sigma^{(0)}(z) + D^{(exp,\gamma) \otimes 2}_{\infty}(z) \ \sigma^{(0)}(z), 
\label{exp}
\end{equation}

\vspace{-4pt}
\noindent where $D^{(exp,\gamma) \otimes 2}_{\infty}$ is the full exponentiated $D_{ee}^{(exp, \gamma)} \otimes D_{ee}^{(exp,\gamma)}$ given by Equation~\eqref{D2beta}. Function $D^{(exp,\gamma) \otimes 2}_{ser}$ is obtained from $D^{(exp,\gamma) \otimes 2}_{\infty}$ by its expansion in series in $\alpha$ and $L$ and keeping only contributions of those orders which are present in $\sigma^{pert}$, namely
the leading and next-to-leading logarithmic contributions up to $\mathcal{O}(\alpha^5L^5)$ and $\mathcal{O}(\alpha^4L^3)$, respectively, together with 
$\mathcal{O}(\alpha^2L^0)$ terms. Let us remind that $\sigma^{pert}$ is the result of our perturbative calculation~\cite{Arbuzov:2024tac}:
\vspace{-4pt}
\begin{eqnarray} \label{sigpert}
&& \hspace{2cm}\sigma^{pert} (z) = D_{e e} \otimes D_{\bar{e}  \bar{e} } \otimes \left(\sigma_{e \bar{e} }^{(0)}(z) + \frac{\alpha}{2 \pi} \overline{\sigma}_{e \bar{e} }^{(1)} (z) + \left( \frac{\alpha}{2 \pi} \right)^2\overline{\sigma}_{e \bar{e} }^{(2)} (z) \right)     \nonumber \\
&& +  2  D_{\gamma e} \!  \otimes  \! D_{\bar{e} \bar{e}} \! \otimes \! \left( \! \frac{\alpha}{2 \pi}\sigma_{e \gamma}^{(0)}(z) + \left( \!  \frac{\alpha}{2 \pi} \! \right)^2 \overline{\sigma}_{e \gamma}^{(1)} (z) \! \right)  + D_{\bar{e} e} \! \otimes \! D_{e \bar{e}} \! \otimes \! \left( \! \frac{\alpha}{2 \pi} \sigma^{(0)}_{e \bar{e} }(z) + \left( \! \frac{\alpha}{2 \pi} \! \right)^2 \overline{\sigma}_{e \bar{e} }^{(1)} (z) \! \right).
\end{eqnarray}

\vspace{-4pt}
\noindent Here $D_{ab}$ are structure functions obtained by iterative solution of the DGLAP equation. 
In this way, we perform matching of the perturbative results with the exponentiated ones
and avoid double counting.
Despite the fact that we have explicit perturbative results only in the LL and NLL approximations and the one $\mathcal{O}(\alpha^2L^0)$ next-to-next-to-leading correction ($h_{20}$), we take into account in $D^{(exp,\gamma) \otimes 2}_{ser}$  all terms up to $\mathcal{O}(\alpha^5L^0)$ appearing in the expansion. All these higher-order corrections are present in $D^{(exp,\gamma) \otimes 2}_{\infty}$, and then cancel each other when we calculate an improved cross section according to Equation~(\ref{exp}). Note that in $D^{(exp,\gamma) \otimes 2}_{ser}(z) \ \sigma^{(0)}(z)$ and $D^{(exp) \otimes 2}_{\infty}(z) \ \sigma^{(0)}(z)$ there is no $\overline{\sigma}^{(1)}$ because all soft and virtual contributions are included into $D^{(exp) \otimes 2}$ and because the exponentiated expressions are applicable at $z \rightarrow 1$ but  $\overline{\sigma}^{(1)}$ tends to zero in this area. To obtain numerical value of the improved cross-section we have to integrate it from some $z_{min}$, defined by experimental conditions, to $1$.

We can estimate the $h_{55}$ pure photonic correction the following way:
\vspace{-4pt}
\begin{equation} \label{h55}
h_{55, \gamma}^{exp}= \left( \ \int  \limits_{z_{min}}^{1} dz D^{(exp, \gamma) \otimes 2}_{ser,55} (z) \ \sigma^{(0)}(sz)  \right)/ \sigma^{(0)} (s) \cdot 100 \%,
\end{equation}

\vspace{-6pt}
\noindent where $D^{(exp, \gamma) \otimes 2}_{ser,55} (z)$ is the part of the expansion that consists of only $\mathcal{O} (\alpha^5 L^5)$ terms.
The values of $h_{55, \gamma}^{exp}$ are presented in percent with respect to Born cross section. Numerical values of the full pure photonic $h_{55}$ and approximated $h_{55, \gamma}^{exp}$ corrections are close to each other, as can be seen from Table~\ref{tab:app1}, which justifies application of the exponentiation for estimates of further higher-order contributions.

We can also estimate the residual “tail” consisting of higher-order contributions provided by the exponentiated result 
beyond the perturbative one:
\vspace{-4pt}
\begin{equation} \label{tail}
\Delta^{D \! D\sigma}_{\infty, \gamma} = \left( \ \int \limits_{z_{min}}^{1} dz D^{(exp, \gamma) \otimes 2}_{\infty} (z) \ \sigma^{(0)}(sz)  - \int \limits_{z_{min}}^{1} dz D^{(exp, \gamma) \otimes 2}_{ser} (z) \ \sigma^{(0)}(sz) \right)/ \sigma^{(0)} (s) \cdot 100 \%.
\end{equation}

\vspace{-4pt}
\noindent We can also estimate pair corrections. Only non-singlet (NS) pair corrections can be exponentiated since the singlet ones are not enhanced at $z\to 1$. Pair contributions come from the running $\alpha$ and the splitting functions. The space-like NLO splitting function of the electron in electron type can be divided into four parts corresponding to pure photonic ($\gamma$) contributions, non-singlet pair ($NS$) ones, singlet pair ($S$) ones, and the interference ($int$) of the latter two ones:
\vspace{-4pt}
\begin{eqnarray}
&& P_{ee}^{(1) \gamma} = 2 \left( \frac{1+x^2}{1-x} \left[ - \ln x \ln (1-x) + \frac{1}{2} \ln^2 x + \mathrm{Li}_2 (1-x)\right] + \ln x + \frac{3}{2} - x + \frac{1}{4} (1+x) \ln^2 x\right) \! ,\\
&& P_{ee}^{(1) NS} = 2 \left( - \frac{1}{3} \frac{1+x^2}{1-x} - \frac{2}{3} (1-x)\right) \! ,\\
&& P_{ee}^{(1) S} = \frac{20}{9 x} -2 +6x - \frac{56}{9} x^2 + (1+5x+\frac{8}{3}x^2) \ln x - (1+x) \ln^2 x,\\
&& P_{ee}^{(1) int} = 2 \left( \frac{1+x^2}{1+x} \left( - \frac{1}{2} \ln^2 x\, \mathrm{Li}_2 (1-x) - \frac{3}{4} \ln x\right) - \frac{7}{4} (1+x) \ln x - 4 + \frac{7}{2} x\right).
\end{eqnarray}

\vspace{-4pt}
In the work \cite{Catani:1989ne} exponentiation of QCD corrections for Drell-Yan and deep inelastic scattering (DIS)
% we added a full form
 processes was discussed taking into account pair corrections. An exponentiation of non-singlet pair and photonic corrections can also be constructed as a sum of two “exponents”.
Here we suggest to perform a joint exponentiation of pure photonic and non-singlet pair corrections in electron PDF. We use the same structure as the expression for pure photonic corrections but modify the parameter $\beta$ as follows:
\vspace{-4pt}
\begin{eqnarray}
D^{(exp,\gamma + NS)}_{ee}(z;\tilde{\beta}) &=& \frac{\tilde{\beta}}{2} (1-z)^{\frac{\tilde{\beta}}{2} - 1} \exp \left( \frac{3}{4} \frac{\tilde{\beta}}{2} \right) \frac{\exp (-\gamma_E \tilde{\beta} /2)}{\Gamma(1+ \tilde{\beta}/2)},
\nonumber \\ \label{DgammaNS}
\tilde{\beta} &\equiv& \frac{2\alpha}{\pi}(L - 1)\left(1+ \frac{\alpha}{6 \pi} \left( L - \frac{7}{3}\right)\right).
\end{eqnarray}

\vspace{-4pt}
\noindent The number $\displaystyle\frac{7}{3}$ was obtained from the comparison of the expanded exponent with the perturbative expression. This coefficient was adjusted to make the terms proportional to $1/(1-z)$ in the contribution of the order $\alpha^2 L^1$ being equal.
The $\Delta$-part of this expression is constructed analogously to Equation~\eqref{dexp}:
\vspace{-2pt}
\begin{equation} \label{dexpNS}
D_{ee}^{(exp, \gamma+NS) \Delta} = \exp \left( \frac{\tilde{\beta}}{2} \ln \Delta  +\frac{3 \tilde{\beta}}{8}\right) \frac{\exp (-\gamma_E \tilde{\beta} /2)}{\Gamma(1+\tilde{\beta}/2)}.
\end{equation}

\begin{figure}[H]
    \centering
    \includegraphics[width=\linewidth]{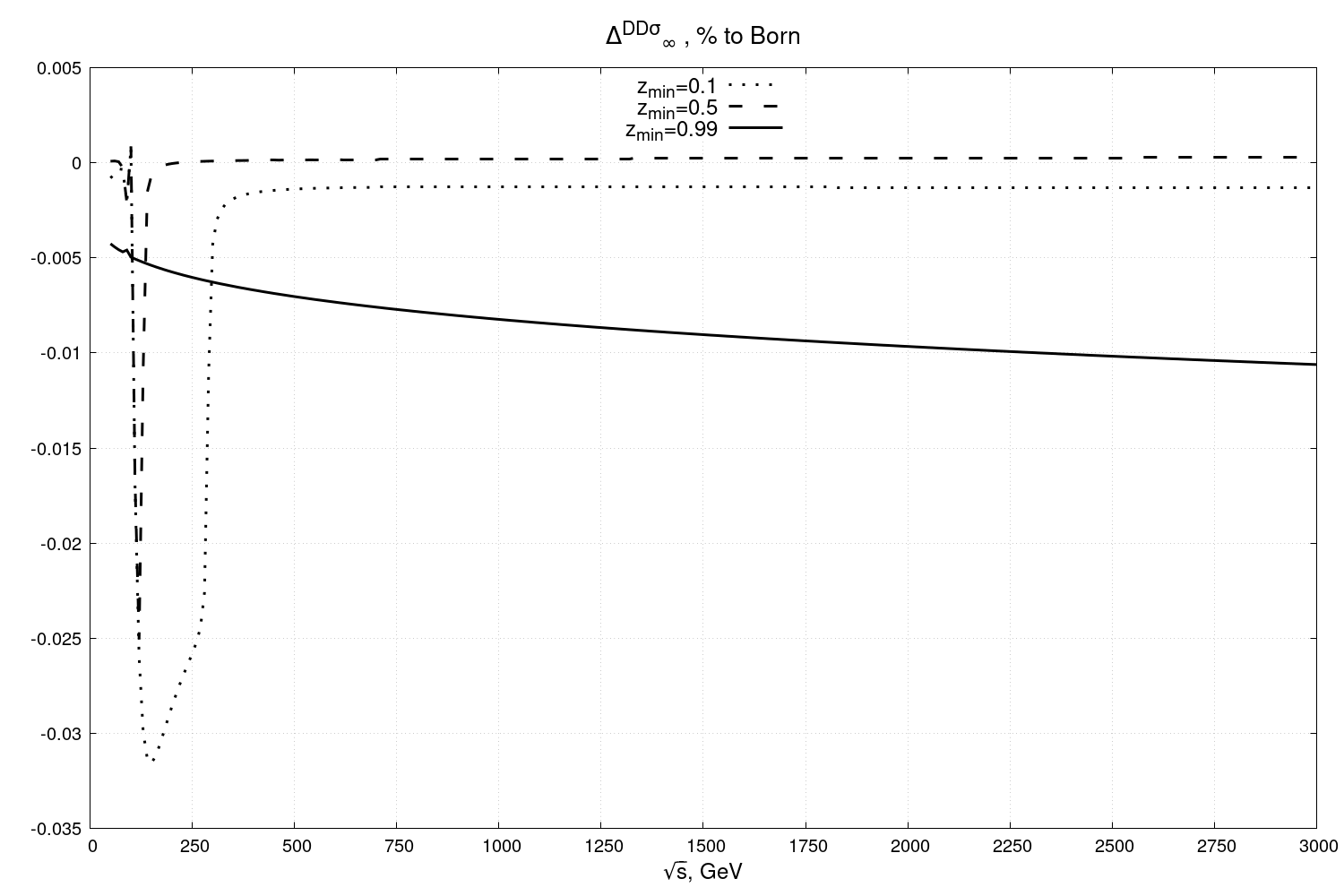}
     \captionsetup{margin={+0.3cm,0cm}}
    \caption{$\Delta^{D \! D\sigma}_{\infty, \gamma+NS}$ in percent \textit{vs.} center-of-mass energy for three different values of $z_{min}$.}
    \label{fig:tails}
\end{figure}

\vspace{-4pt}
\noindent Note that Equation~(\ref{DgammaNS}) takes into account pure photonic contributions, pure non-singlet pair ones, and mixed (photonic + non-singlet pair) corrections. 
We can estimate photonic plus non-singlet pair corrections in the same way as in Equation~(\ref{h55}), and 
$\Delta^{D \! D\sigma}_{\infty, \gamma+NS}$ the same way as in Equation~\eqref{tail}.  In Table~\ref{tab:app1}, they are presented in the columns $h^{exp}_{55}$ and $\Delta^{D \! D\sigma}_{\infty}$ correspondingly, pure photonic case is shown in the rows marked “$\gamma$”, 
and photonic plus non-singlet pair case is shown in the rows marked “Full”. In Figure~\ref{fig:tails} we show the dependence of the contribution of the exponentiation “tail” $\Delta^{D \! D\sigma}_{\infty, \gamma+NS}$ on the center-of-mass energy for three different values of the cut value $z_{min}$. One can see that the effect of exponentiation is can be relevant numerically if the precision level of $10^{-4}$ in the total cross section measurement will be achieved. The relatively wide deep in the case of $z_{min}=0.1$ corresponds to the effect of the radiative return to the $Z$ resonance.

\vspace{-10pt}

\vspace{-18pt}
\section{Scheme dependence}

Despite the fact that the sum of all corrections of particular order in $\alpha$ doesn't depend neither on the factorization scale value nor on the subtraction scheme choice, the expressions for individual contributions \textls[+12]{are scheme-dependent. In the $\overline{\mathrm{MS}}$ scheme we use the following initial conditions for the \mbox{evolution equation}:}
\vspace{-8pt}
\begin{eqnarray} \label{ddd1}
&& d_{ee}^{(1)} (x)=  \left[\frac{1 +x^2}{1-x}(- 1 - 2 \ln(1-x) )\right]_+, \\
&& d_{\gamma e}^{(1)}(x) = -\frac{1 + (1-x)^2}{ x}(2 \ln x + 1), \\
&& \hspace{+2cm}d_{e \gamma}^{(1)}(x) = 0,
\end{eqnarray}

\vspace{-6pt}
\noindent and the expression for the relevant NLO splitting function reads
\vspace{-4pt}
\begin{eqnarray}
&& P_{ee}^{(1)\Theta}(x) = -\frac{56 x^2}{9}+\left(\frac{8 x^2}{3}+\frac{11 x}{3}-\frac{13}{3 (1-x)}+\frac{5}{3}\right) \ln (x) +\frac{37 x}{3}+\frac{20}{9 x} -\left(\frac{3
   x}{2}+\frac{3}{2}\right) \ln ^2(x) \nonumber \\
&&  \hspace{+2.5cm}\quad + \left((2 x+2) \ln (1-x)  -\frac{4 \ln
   (1-x)}{1-x}\right) \ln (x)-\frac{25}{3}, \nonumber \\
&& \hspace{+4.8cm} P_{ee}^{(1) \Delta} =\frac{15}{8} - \frac{13}{3} \zeta_2 + 6 \zeta_3.
\end{eqnarray}

\vspace{-6pt}
\noindent In Equation~(\ref{ddd1}) the subscript “$+$” means the standard plus prescription. 
If we compare an expression for $D_{ee}(x)$ obtained from iterative solution of the evolution equation in the $\overline{\mathrm{MS}}$ scheme with the $D^{(exp, \gamma)}_{ser}$, we can see that the $\overline{\mathrm{MS}}$ scheme generates higher powers of $\ln(1-z)$ than exponentiation due to the presence of $\frac{1 +x^2}{1-x} \ln(1-x)$ in $d_{ee}^{(1)}$. In order-by-order calculations those singular at $z\to1$ extra terms (they do not correspond to any soft or collinear singularities) cancel out with the corresponding terms from convolution of the leading-order structure function with the
next-to-leading order partonic cross section $\overline{\sigma}^{(1)}$. But in the exponentiated $D^{(exp \otimes 2)}$ multiplied by $\sigma^{(0)}$ they don't cancel out.

We suggest another factorization scheme, where
\vspace{-6pt}
\begin{equation}
    \tilde{d}^{(1)}_{ee} (x)= - \left[\frac{1+x^2}{1-x}\right]_+. 
    %= - P_{ee}^{(0)} (x). 
\end{equation}

\vspace{-8pt}
\noindent So, we change $d_{ee}^{(1)}$ and derive the corresponding expressions for $\widetilde{\sigma}^{(1)}$ and $P_{ee}^{(1)}$ from matching equalities at
\vspace{-6pt}
 $\mathcal{O}(\alpha^1)$~\cite{Arbuzov:2024tac} and $\mathcal{O}(\alpha^2)$:
\begin{equation} \label{matchingOa1}
\delta^{(1)}_{\bar{e} e}(sx) = \bar{\delta}^{(1)}_{\bar{e} e}(sx) 
+ 2 \frac{\alpha}{2\pi}\left[ P^{(0)}_{ee}(y) L + d_{ee}^{(1)}(y)\right] 
+ \mathcal{O}\left(\frac{m_e^2}{s}\right),
\end{equation}

\vspace{-6pt}
\noindent where $\bar{\delta}_{\bar{e} e}^{(1)} (sx)\equiv \frac{\overline{\sigma}_{\bar{e} e}^{(1)} (sx)}{\sigma_{\bar{e} e}^{(0)} (sx)}$ and $\delta_{\bar{e} e}^{(1)} (sx)\equiv \frac{\sigma_{\bar{e} e}^{(1)} (sx)}{\sigma_{\bar{e} e}^{(0)} (sx)}$, and
\vspace{-4pt}
\begin{eqnarray}
&& \delta^{(2)}_{\bar{e} e}(sx) = \left( \frac{\alpha}{2\pi}\right)^2 L^2 \left[ P_{e \gamma}^{(0)} \otimes P_{\gamma e}^{(0)} + \frac{2}{3} P_{ee}^{(0)} + 2 P_{ee}^{(0)\otimes 2}\right] + \left( \frac{\alpha}{2\pi}\right)^2 L \left[ 2  \bar{\delta}_{e \gamma}^{(0)} \otimes P_{\gamma e}^{(0)} + \frac{2}{3}\bar{\delta}_{\bar{e} e}^{(1)} \right. \nonumber \\
&& \hspace{+0.5cm}\quad \left.+ 2 \bar{\delta}_{\bar{e} e}^{(1)} \otimes P_{ee}^{(0)} + 2 d_{\gamma e}^{(1)} \otimes  P_{e \gamma}^{(0)} +2 P_{ee}^{(1)} - \frac{20}{9} P_{ee}^{(0)} + 4 P_{ee}^{(0)} \otimes d_{ee}^{(1)}\right] + \left(\frac{\alpha}{2\pi}\right)^2 c_{20}.
\end{eqnarray}
 
 \vspace{-4pt}
\noindent We omit the arguments on the right-hand side for brevity. Functions $\overline{\sigma}^{(2)}$ and $d_{ee}^{(2)}$, which will appear in $c_{20}$, must change due to change of $d_{ee}^{(1)}$, $P_{ee}^{(1)}$ and $\overline{\sigma}^{(1)}$. We assume that $P_{e \gamma}^{(1)}$, $P_{\gamma e}^{(1)}$, and $d_{\gamma e}^{(0)}$ don't change because they contain no unphysical logarithms. From this we can conclude that it is enough to consider the sum 
\vspace{+6pt}
\begin{equation}
    \frac{2}{3}\bar{\delta}_{\bar{e} e}^{(1)} + 2 \bar{\delta}_{\bar{e} e}^{(1)} \otimes P_{ee}^{(0)}  +2 P_{ee}^{(1)} + 4 P_{ee}^{(0)} \otimes d_{ee}^{(1)},
\end{equation}

\noindent which must not change for matching. In our new scheme, we have
\vspace{-6pt}
\begin{eqnarray}
   &&\hspace{-0.65cm} \widetilde{P}_{ee}^{(1)}(x) = {P}_{ee}^{(1)}(x) + \frac{4}{3} (1+x^2) \frac{\ln(1-x)}{1-x}, \nonumber \\
   && \widetilde{P}_{ee}^{(1) \Delta} = - \frac{11}{12}  + \frac{8}{3} \ln^2 \Delta - \frac{26}{3} \zeta_2, \nonumber \\
    &&\hspace{+0.16cm}\widetilde{\overline{\sigma}}_{ee}^{(1)} (x) = 2 \frac{1+x^2}{1-x} (\ln z - \ln x), \\
    &&\hspace{+1.3cm}\widetilde{\overline{\sigma}}_{ee}^{(1) \Delta} = 4 \zeta_2 - 1.
\end{eqnarray}

\vspace{-8pt}
\noindent \textls[-16]{In our new scheme $\widetilde{\overline{\sigma}}_{ee}^{(1)}$ has its theta part equal to zero at $x=z$, and convolutions become significantly easier. }

\section{Theoretical uncertainty}

We can estimate the following three different contributions to theoretical uncertainty:
\vspace{-6pt}
\begin{eqnarray}
    && \Delta_{LL,exp} =  \frac{h_{55}}{h_{44}}  (h^{(exp)}_{55} - h_{55}), \label{erh66} \\
    &&\hspace{+0.8cm} \Delta_{NLL} = \frac{h_{43}}{h_{32}} h_{43}, \label{h54} \\
    && \hspace{+0.72cm}\Delta_{NNLL} = \frac{h_{20}}{h_{21}} h_{32}. \label{h31}
\end{eqnarray}

\vspace{-4pt}
\noindent Equation \eqref{erh66} estimates the uncertainty in $h_{66}$; Equation \eqref{h54} estimates the unaccounted NLL contribution $h_{54}$, and Equation \eqref{h31} estimates the unaccounted NNLL contribution $h_{31}$.
In the Table \ref{table-uncert} we present the results for calculating uncertainties at the Z-peak where the highest experimental precision 
of the order of $10^{-3}\%$ can be achieved at future colliders in the so-called Tera-Z mode~\cite{Blondel:2018mad}.

	\begin{table}[H] 
    \centering
    \fontsize{10pt}{12pt}\selectfont
        \caption{Theoretical uncertainty in \% at $\sqrt{s}=M_Z$.}%Please include a textual description of Table 1 in the main text, and make sure that this description precedes the table. 
		\begin{tabular}{lccc} 
		\toprule
        ${z}_{\mathbfit{min}}$ & ${\Delta}_{\mathbfit{LL,exp}}$ & ${\Delta}_{\mathbfit{NLL}}$ & ${\Delta}_{\mathbfit{NNLL}}$ \\
        \midrule
        0.1 &  0.00019 & 0.00014 & 0.00036 \\ 
        
        0.5 & 0.00019 & 0.00014 & 0.00036 \\
    
        0.9 &  0.00021 & 0.00014 & 0.00035 \\
        \bottomrule
        \end{tabular}
        \label{table-uncert}
    \end{table}

\vspace{-4pt}
Here all the values are in percent with respect to Born cross section, so the current theoretical uncertainty in description of ISR in the total cross section at $Z$-peak can be estimated to be about $4\cdot 10^{-4}\%$. Obviously, calculation of the $h_{31}$ contribution should be the next step to improve the accuracy. 

At the energy $\sqrt{s}=240$ GeV the uncertainty is bigger for $z_{min}=0.1$ because of the radiative return to the resonance, but the experimental requirements are not so strict. Uncertainties in this energy domain will be considered elsewhere.

\section{Conclusion}
%\textbf{\large 4. Conclusion}

So, we constructed matching of the existing perturbative results for the leading and next-to-leading
higher-order ISR corrections to processes of $e^+e^-$-annihilation with simultaneous exponentiation
of certain photonic and pair corrections. The matching allows one to combine the advantages of 
the two approaches and thus to improve the resulting precision. Note that any future perturbative
result can be easily incorporated within the proposed scheme. 

Numerical values of LL and NLL ISR corrections to the cross section of electron-positron 
annihilation in $\mathrm{\overline{MS}}$ scheme were calculated. The calculations will be implemented 
in the ZFITTER program~\cite{Akhundov:2013ons}.
The suggested new subtraction scheme is useful for evaluation of radiative corrections
within the QED structure approach, especially for $z\to 1$. The theoretical uncertainty 
in the description of radiative corrections in $e^+e^-$ annihilation processes should 
still be improved by inclusion of new perturbative calculations, including effects due to the
final state radiation (FSR) and ISR-FSR interference.

The presented numerical results are obtained for inclusive treatment of the ISR effects.
Alternatively, one can apply QED parton showers, which are a kind of exponentiated 
description of ISR in a differential form. If the latter is relevant for a given observable, one should perform matching of the completely differential perturbative results with parton showers. Our work on this task is in progress within the ReneSANCe Monte Carlo event 
generator~\cite{Sadykov:2020any}.

\section*{Data availability statement}
No supplementary or additional data were generated in this study. 

\section*{Declaration of generative AI and AI-assisted technologies}
The authors did not use generative AI or AI-assisted technologies in the writing of this manuscript.

\section*{Acknowledgments}
We are grateful to A.V.~Kotikov for fruitful discussions.

\section*{Appendix}
\renewcommand{\thetable}{A\arabic{table}}  
\setcounter{table}{0}

\renewcommand{\thetable}{A\arabic{table}} 

\makeatletter
\def\fnum@table{\tablename~\thetable}  
\makeatother

Here we present Table~\ref{tab:app1} with numerical results for particular contributions.

\newpage
{\fontsize{8pt}{8.8pt}\selectfont
\setlength{\tabcolsep}{3.2pt} % 压缩列间隙适配13列
\begin{longtable}{lcccccccccccc}
\caption{Numerical values of radiative corrections, $\%$.} \label{tab:app1} \\
\toprule
& $\pmb{h_{11}}$ & $\pmb{h_{10}}$ & $\pmb{h_{22}}$ & $\pmb{h_{21}}$ & $\pmb{h_{20}}$ & $\pmb{h_{33}}$ &$\pmb{h_{32}}$ & $\pmb{h_{44}}$ & $\pmb{h_{43}}$ & $\pmb{h_{55}}$ & $\pmb{h_{55}^{\mathrm{exp}}}$ &$\pmb{\Delta^{D\,D\sigma}_\infty}$ \\
\midrule
\endfirsthead

\multicolumn{13}{c}{\fontsize{12pt}{17pt}\selectfont{\textbf{\tablename\ \thetable{.}} \textit{Cont.}}} \\[12pt]
\toprule
& $\pmb{h_{11}}$ & $\pmb{h_{10}}$ & $\pmb{h_{22}}$ & $\pmb{h_{21}}$ & $\pmb{h_{20}}$ & $\pmb{h_{33}}$ &$\pmb{h_{32}}$ & $\pmb{h_{44}}$ & $\pmb{h_{43}}$ & $\pmb{h_{55}}$ & $\pmb{h_{55}^{\mathrm{exp}}}$ &$\pmb{\Delta^{D\,D\sigma}_\infty}$\\
\midrule
\endhead

\midrule
\multicolumn{13}{r}{{\footnotesize}} \\
\endfoot

\bottomrule
\endlastfoot

        \multicolumn{13}{c}{$\pmb{\sqrt{s}=M_Z - 1}$} \\
		\midrule
		\multicolumn{13}{c}{$z_{min}=0.1$}\\
		\midrule 
		  $\gamma$ & $-$37.62744 & 2.20532  & 7.28672
		& $-$0.82576 & 0.02063
		& $-$0.95300 & 0.15752 & 0.09393 & $-$0.02019 & $-$0.00737 & $-$0.00714
        & $-$0.00200\\
		
		  Pairs & 0 & 0
		& $-$0.35072
		& 0.20364 & $-$0.03280
		& 0.13186 & $-$0.08364 & $-$0.02446 & 0.01642 & 0.00297 & - & -\\
		
		  Full  & $-$37.62744 & 2.20532  & 6.93600
		& $-$0.62292 & $-$0.01217
		& $-$0.82114 & 0.07388 & 0.06947 & $-$0.00377 & $-$0.00441 & $-$0.00385 & $-$0.00216\\
        \midrule
		\multicolumn{13}{c}{$z_{min}=0.5$}\\
		\midrule 
		  $\gamma$ & $-$37.66044 & 2.2067  & 7.28480
		& $-$0.82656 & 0.02065
		& $-$0.95285 & 0.15746 & 0.09393 & $-$0.02019 & $-$0.00737 & $-$0.00715 & $-$0.00200\\
		
		  Pairs & 0 &0   & $-$0.35204
		& 0.2038 & $-$0.03277
		& 0.13185 & $-$0.08363 & $-$0.002446 & 0.01642 & 0.00297 & - & - \\
		
		  Full  & $-$37.66044 & 2.2067 
		& 6.93272
		& $-$0.62276 & $-$0.01213
		& $-$0.82101 & 0.07383 & 0.06947 & $-$0.00377 & $-$0.00440 & $-$0.00386 & $-$0.00216 \\
		\midrule
		\multicolumn{13}{c}{$z_{min}=0.9$}\\
		\midrule 
		  $\gamma$ & $-$38.07502
		& 2.22386  & 7.2852
		& $-$0.82684 & 0.02085
		& $-$0.95183 & 0.15758 & 0.09337 & $-$0.02007 & $-$0.00732 & $-$0.00706 & $-$0.00202\\
		
		  Pairs & 0 &  0 & $-$0.35244
		& 0.20392 & $-$0.03288
		& 0.13271 & $-$0.08393 & $-$0.02442 & 0.01637 & 0.00295 & - & -\\
		
		  Full & $-$38.07502
		& 2.22386
		& 6.93272
		& $-$0.62276 & $-$0.01203
		& $-$0.81912 & 0.07365 & 0.06895 & $-$0.00370 & $-$0.00438 & $-$0.00382 &  $-$0.00216 \\
		\midrule 
		\multicolumn{13}{c}{$z_{min}=0.99$}\\
		\midrule 
		  $\gamma$ & $-$46.70978
		& 2.59612  & 10.41804
		& $-$1.14888 & $-$0.02851
		& $-$1.40697 & 0.23324 & 0.12309 & $-$0.02747 & $-$0.00622 & $-$ 0.00597 & $-$0.00462 \\
		
		  Pairs & 0 & 0  & $-$0.44028
		& 0.2706 & $-$0.04487
		& 0.18941 & $-$0.12097 & $-$0.03622 & 0.02306 & 0.00380 & - &- \\
		
		  Full & $-$46.70978
		& 2.59612
		& 9.9778
		& $-$0.87832 & $-$0.01636
		& $-$1.21756 & 0.11228 & 0.008687 & $-$0.00441 & $-$0.00242 & $-$0.00169  & $-$0.00462 \\
		
        \midrule
    \multicolumn{13}{c}{$\pmb{\sqrt{s}=M_Z + 1}$}\\
   \midrule
		\multicolumn{13}{c}{$z_{min}=0.1$}\\
		\midrule
		  $\gamma$ & $-$10.85409
		& 1.09644 & $-$3.20789
		& 0.21972 & $-$0.00441
		& 1.04577 & $-$0.15565 & $-$0.14951 & 0.03139 & 0.01349 & 0.01371 & 0.00090\\
		
		  Pairs & 0 & 0  & $-$0.10032
		& $-$0.00757 & 0.01180
		& $-$0.06152 & 0.06022 & 0.02836 & $-$0.02194 & $-$0.00498 & - & - \\
		
		  Full & $-$10.85409
		& 1.09644
		& $-$3.30821
		& 0.21215 & 0.00738
		& 0.98425 & $-$0.09543 & $-$0.12115 & 0.00945 & 0.00850 & 0.00832 & 0.00238 \\
		\midrule 
		\multicolumn{13}{c}{$z_{min}=0.5$}\\
		\midrule 
		  $\gamma$ & $-$10.88572
		& 1.09775  & $-$3.20974
		& 0.21972 & $-$0.00440
		& 1.04590 & $-$0.15571 & $-$0.14950 & 0.03139 & 0.01349 & 0.01370 &  0.00090 \\
		
		  Pairs &  0& 0  & $-$0.10159
		& $-$0.00742 & 0.01182
		& $-$0.06153 & 0.06023 & 0.02836 & $-$0.02194 & $-$0.00498 & - & - \\
		
		  Full & $-$10.88572
		& 1.09775
		& $-$3.31133
		& 0.21230 & 0.00742
		& 0.98438 & $-$0.09548 & $-$0.12114 & 0.00945 & 0.00850 & 0.00832 & 0.00238 \\
		\midrule
		\multicolumn{13}{c}{$z_{min}=0.9$}\\
		\midrule 
		  $\gamma$ &  $-$11.55686
		& 1.12548  & $-$3.12720
		& 0.20817 & $-$0.00407
		& 1.04707 & $-$0.15539 & $-$0.15045 & 0.03158 & 0.01357 & 0.01381 & 0.00087 \\
		
		  Pairs & 0 & 0   & $-$0.10808
		& $-$0.00438 & 0.01163
		& $-$0.06003 & 0.05969 & 0.02842 & $-$0.02203 & $-$0.00502 & - & -\\
		
		  Full & $-$11.55686
		& 1.12548
		& $-$3.23528
		& 0.20379 & 0.00756
		& 0.98704 & $-$0.09571 & $-$0.12206 & 0.00905 & 0.00855 &  0.00838 & 0.00238 \\
		\midrule
		\multicolumn{13}{c}{$z_{min}=0.99$}\\
		\midrule 
		  $\gamma$ & $-$39.28805
		& 2.27114  & 6.15316
		& $-$0.74405 & 0.01908
		& $-$0.26952 & 0.06537 & $-$0.07252 & 0.01158 & 0.01794 & 0.01828 & $-$0.00705\\
		
		  Pairs & 0 & 0  & $-$0.36817
		& 0.19250 & $-$0.02386
		& 0.11072 & $-$0.04958 & $-$0.00566 & $-$0.00330 & $-$0.00285&  - &-\\
		
		Full & $-$39.28805
		& 2.27114
		& 5.78498
		& $-$0.55155 & $-$0.00477
		& $-$0.15880 & 0.01579 & $-$0.07818 & 0.00828 & 0.01509 & 0.01551 & $-$0.00493\\
        \midrule
        \multicolumn{13}{c}{$\pmb{\sqrt{s}=M_Z}$}\\
        		\midrule
		\multicolumn{13}{c}{$z_{min}=0.1$}\\
		\midrule
		  $\gamma$ & $-$32.73654
		& 2.00167 & 4.88428
		& $-$0.59516 & 0.01521
		& $-$0.37760 & 0.07104 & 0.00344 & $-$0.00185 & 0.00315 & 0.00341 & -0.00270\\
		
		  Pairs & 0  & 0  & $-$0.30571
		& 0.15847 & $-$0.02152
		& 0.08758 & $-$0.04604 & $-$0.00909 & 0.00375 & $-$0.00008 & - & - \\
		
		  Full &$-$32.73654
		& 2.00167
		& 4.57858
		& $-$0.43669 & $-$0.00631
		& $-$0.29002 & 0.02500 & $-$0.00564 & 0.00190 & 0.00307 & 0.00344 & $-$0.00194\\
		\midrule
		\multicolumn{13}{c}{$z_{min}=0.5$}\\
		\midrule 
		  $\gamma$ & $-$32.75622
		& 2.00249  & 4.88313
		& $-$0.59516 & 0.01521
		& $-$0.37751 & 0.07101 & 0.00345 & $-$0.00185 & 0.00315 & 0.00341 & $-$0.00270\\
		
		  Pairs & 0& 0  & $-$0.30650
		& 0.15857 & $-$0.02150
		& 0.08757 & $-$0.04603 & 0.00908 & 0.00375 & $-$0.00008 & - & -\\
		
		  Full & $-$32.75622
		& 2.00249
		& 4.57663
		& $-$0.43660 & $-$0.00629
		& $-$0.28994 & 0.02497 & $-$0.00564 & 0.00190 & 0.00307 & 0.00344 & $-$0.00194\\
		\midrule
		\multicolumn{13}{c}{$z_{min}=0.9$}\\
		\midrule
		  $\gamma$  & $-$33.0710
		& 2.01547  & 4.92033
		& $-$0.60045 & 0.01537
		& $-$0.37684 & 0.07112 & 0.00301 & $-$0.00176 & 0.00319 & 0.00345 & $-$0.00271\\
		
 		 Pairs  & 0& 0  & $-$0.30954
		& 0.15997 & $-$0.02157
		& 0.08824 & $-$0.04627 & 0.00905 & 0.00371 & $-$0.00009 & - & -\\
		 
		  Full & $-$33.0710
		& 2.01547
		& 4.61079
		& $-$0.44048 & $-$0.00622
		& $-$0.28860 & 0.02485 & $-$0.00604 & 0.00195 & 0.00310& 0.00347 & $-$0.00194\\

     \multicolumn{13}{c}{$\pmb{\sqrt{s}=160}$ \textbf{GeV}}\\   
     		\midrule
	\multicolumn{13}{c}{$z_{min}=0.1$}\\
		\midrule
		  $\gamma$ & 335.79956 &  $-$12.62015 & 17.14787
		& 0.64666 & $-$0.11553
		& $-$1.82233 & 0.57813 &  $-$0.03476 & $-$0.01018 & 0.00562 & 0.01695 & $-$0.03213\\
		 
		  Pairs & 0 & 0
		& 8.35391
		& $-$1.84375 & $-$0.14402
		&  0.13309 & $-$0.09339 & $-$0.06055 & 0.03570 & $-$0.00164 & - & -\\
		 
		Full & 335.79956&  $-$12.62015 &  25.50178
            &  $-$1.19709 & $-$0.25954
		&$-$1.68924& 0.48483& $-$0.09531 &  0.02552 & 0.00397 & 0.02057  & $-$0.00312 \\
        \midrule
		\multicolumn{13}{c}{$z_{min}=0.5$}\\
		\midrule
		  $\gamma$ & 9.81624 & 0.26017  & $-$1.28618
		& 0.20722 & $-$0.00761
		& $-$0.00845 & $-$0.00714 & 0.00530 & $-$0.00153 & $-$0.00022  & $-$0.00097 & 0.00032\\
		 
		  Pairs & 0 & 0
		& 0.13182
		& $-$0.05573 & $-$0.00427
		& $-$0.02757 & 0.01162 & $-$0.00171 & 0.00046 & 0.00020 & - & -\\
		 
		  Full & 9.81624 & 0.26017  & $-$1.15435
		& 0.15149 & $-$0.01188
		& $-$0.03602 & 0.00447 & 0.00359 & $-$0.00107 & 0.00002 & $-$0.00018 & $-$0.00039\\
		\midrule
		\multicolumn{13}{c}{$z_{min}=0.9$}\\
			\midrule 
		  $\gamma$ & $-$17.49251 &  1.33920& 0.37598
		& $-$0.11629 & 0.00320
		& 0.17976 & $-$0.02564 & $-$0.02424 & 0.00549 & 0.00143 & 0.00156 & 0.00072\\
		
 		 Pairs &  0 & 0
		& $-$0.17090
		& 0.05814 & $-$0.00496
		& 0.00499 & 0.00377 & 0.00511 & $-$0.00376 & $-$0.00083 & - & -\\
		   
		  Full & $-$17.49251 &  1.33920   & 0.20508
		& $-$0.05815 & $-$0.00177
		& 0.18475 & $-$0.02187 & $-$0.01912 &  0.00174&  0.00058 & 0.00060 & 0.000103\\
            \midrule
    \multicolumn{13}{c}{$\pmb{\sqrt{s}=240}$ \textbf{GeV}}\\
   \midrule	 
    	\multicolumn{13}{c}{$z_{min}=0.1$}\\
		 \midrule
		  $\gamma$ &  324.14638&  $-$11.76206 & 27.12979
		& $-$0.99088 &  $-$0.08026
		& $-$0.62423 & 0.61404 & $-$0.06641 & 0.01487 & 0.00247 & 0.02316 & $-$0.03299\\
		
		  Pairs & 0 & 0
		&  25.33367
		& $-$1.47717 & $-$0.44478
		&  $-$0.02167& $-$0.03074 & $-$0.01918 & 0.04599& 0.00012 & - & -\\
		
		  Full & 324.14638&  $-$11.76206 &  52.46346
            &  $-$2.46805 & $-$0.52504
		& $-$0.63690 &0.58330 & $-$0.08559 &  0.06086 & 0.00259 & 0.01950 & $-$0.02631 \\
		\midrule
		\multicolumn{13}{c}{$z_{min}=0.5$}\\
		\midrule
		  {$\gamma$} & 3.45872 & 0.51562  & $-$1.05495
		& 0.14324 & $-$0.00510
		& 0.02506 & $-$0.01176 & 0.00273 & $-$0.00067& $-$0.00020 & $-$0.00089 & 0.00059\\
		
 		 Pairs &  0 & 0
		& 0.05956
		& $-$0.02900 &  $-$0.00412
		& $-$0.02361 & $-$0.01040 & 0.00046 & $-$0.00017 & 0.00013 & - & -\\
		
		  Full & 3.45872 & 0.51562  & $-$0.99539
		& 0,11424 & $-$0.00923
		& 0,00145 & $-$0,00136& 0.00319 & $-$0,00084 & $-$0,00007 & $-$0.00005 & 0.00003\\
		\midrule
		\multicolumn{13}{c}{$z_{min}=0.9$ }\\
		\midrule 
		  $\gamma$ & $-$18.60745 & 1.36043 & 0.56697 
		& $-$0.13653 & 0.00358
		& 0.17533 & $-$0.02372 & $-$0.02589  & 0.00568  & 0.00161 & 0.00179 & 0.00063\\
		 
		  Pairs & 0 & 0
		& $-$0.18765
		& 0.06353 & $-$0.00549
		& 0.00878 & 0.00216 &  0.00551 & $-$0.00380  & $-$0.00092 & - & -\\
		
		  Full & $-$18.60745 & 1.36043   &  0.37931
		&  $-$0.07300 & $-$0.00191
		& 0.18411 & $-$0.02157  & $-$0.02038 &  0.00188 &  0.00069 & 0.00073&  0.00098\\
		 
        \midrule
        \multicolumn{13}{c}{$\pmb{\sqrt{s}=3}$ \textbf{TeV}} \\
         \midrule		
    	\multicolumn{13}{c}{$z_{min}=0.1$}\\
		\midrule 
		 $\gamma$ & 19.53776 &  0.02118 &  0.03886
		& 0.14215 & $-$0.00856
		& $-$0.06571  & 0.02188  & 0.00028  & $-$0.00096 & 0.0001 & 0.00014 & $-$0.00066 \\
		 
		  Pairs & 0 & 0
		& 1.03833  
		& $-$0.11012  & $-$0.01629
		& $-$0.02899  & 0.00887  & $-$0.00263  & 0.00276  & $-$0.01574 & - & \\
		
		  Full & 19.53776 &  0.02118  &   
          1.07719  &  0.03203  & $-$0.02485
		& $-$0.0947  & 0.03075 & 0.00235 & 0.00180  & $-$0.01564 & 0.00148 & $-$0.00133\\
		\midrule
		\multicolumn{13}{c}{$z_{min}=0.5$}\\
		\midrule 
		  $\gamma$ &  2.22920 &  0.57652 & $-$1.34517 
		& 0.14759  & $-$0.00447
		&  0.04977 & $-$0.01709 & 0.00413  &  $-$0.00078 & $-$0.00043 & $-$0.00195 & 0.00113\\
		
		  Pairs &  0 & 0
		& 0.05826
		& $-$0.02631 & $-$0.00422
		& $-$0.03629 & 0.01365 & 0.00126  & $-$0.00047  & 0.00024 & - & -\\
		
		Full & 2.22920 &  0.57652  &  $-$1.28691
		&  0.12128 & $-$0.00869
		& 0.01349  & $-$0.00344 &  0.00539 &  $-$0.00125 & $-$0.00018 & $-$0.00073 & 0.00027\\
		\midrule
		\multicolumn{13}{c}{$z_{min}=0.9$ }\\
		\midrule 
		  $\gamma$ & $-$22.44302  &  1.36803  & 0.89303  
		&  $-$0.17003 & 0.00371
		& 0.28417  & $-$0.03192   & $-$0.05129   & 0.00942  & 0.00384 & 0.00428 & 0.00052\\
		
		  Pairs & 0 & 0
		&  $-$0.27012
		&   0.07734& $-$0.00568
		& 0.01695  &  0.00211 &  0.01072  &  $-$0.00624  &  $-$0.00218 & -  & -\\
		
		  Full & $-$22.44302  &  1.36803     
		&  0.06229  & $-$0.0927
		&  $-$0.00197 &  0.30112 &  $-$0.02981 & $-$0.04056 &0.00318 &   0.00166 & 0.00179 & 0.00114 \\[-10pt]
%    \label{tab:app1}
\end{longtable}
}

%\section*{References}

%\bibliographystyle{aiasbst}
\bibliographystyle{elsarticle-num}
\bibliography{Exponentiation_arxiv_v2}

@inproceedings{Blondel:2018mad,
    author = "Blondel, A. and others",
    title = "{Standard model theory for the FCC-ee Tera-Z stage}",
    booktitle = "{Mini Workshop on Precision EW and QCD Calculations for the FCC Studies : Methods and Techniques}",
    eprint = "1809.01830",
    archivePrefix = "arXiv",
    primaryClass = "hep-ph",
    reportNumber = "CERN-2019-003, BU-HEPP-18-04, CERN-TH-2018-145, IFJ-PAN-IV-2018-09, KW 18-003, MITP/18-052, MPP-2018-143, SI-HEP-2018-21",
    publisher = "CERN",
    address = "Geneva",
    series = "CERN Yellow Reports: Monographs",
    volume = "3/2019",
    month = "9",
    year = "2018"
}

@article{Sadykov:2020any,
    author = {Sadykov, Renat and Yermolchyk, Vitaly},
    title = {\textls[-24]{Polarized NLO EW $e^+e^-$ cross section calculations with ReneSANCe-v1.0.0}},
    journal = {Comput. Phys. Commun.},
    volume = {256},
    pages = {107445},
    year = {2020}
}

@article{Golonka:2005pn,
    author = {Golonka, Piotr and Was, Zbigniew},
    title = {PHOTOS Monte Carlo: a precision tool for QED corrections in $Z$ and $W$ decays},
    journal = {Eur. Phys. J. C},
    volume = {45},
    number={1},
    pages = {97--107},
    year = {2006}
}

@article{Ward:2026bco,
    author = {Ward, B. F. L. and Jadach, S. and Placzek, W. and Skrzypek, M. and Was, Z. and Yost, S. and Siodmok, A.},
    title = {\textls[-18]{Recent developments in IR-improved amplitude-based resummation in precision high energy collider physics}},
journal={arXiv},
volume={arXiv:2604.26124},
    year = {2026}
}

@inproceedings{Frixione:2022ofv,
    author = "Frixione, S. and others",
    title = "{Initial state QED radiation aspects for future $e^+e^-$ colliders}",
    booktitle = "{Snowmass 2021}",
    eprint = "2203.12557",
    archivePrefix = "arXiv",
    primaryClass = "hep-ph",
    reportNumber = "FERMILAB-PUB-22-129-T",
    month = "3",
    year = "2022"
}

@article{Jadach:1998jb,
    author = {Jadach, S. and Ward, B. F. L. and Was, Z.},
    title = {Coherent exclusive exponentiation CEEX: the case of the resonant $e^+e^-$ collision},
    journal = {Phys. Lett. B},
    volume = {449},
    pages = {97--108},
    year = {1999}
}

@incollection{Jadach:1989nm,
    author = {Jadach, Stanislaw and Ward, B. F. L.},
    title = {Exclusive exponentiation in the Monte Carlo Yennie-Frautschi-Suura approach},
    booktitle = {\textit{NATO ASI Series}},
    address={1st ed. New York},
    publisher={Springer,~},
    note={\MakeLowercase{pp. 325--340}},
    year = {1989}
}

@article{Arbuzov:2024tac,
    author = {Arbuzov, A. and Voznaya, U.},
    title = {\textls[+18]{Higher-order NLO initial state QED radiative corrections to $e^+e^-$} annihilation revisited},
    journal = {Phys. Rev. D},
    year = {2024},
    volume = {109},
    number = {11},
    pages = {113002},
}

@article{Passarino:2001wy,
    author = {Passarino, Giampiero},
    title = {A Practical approach for exponentiation of QED corrections in arbitrary processes},
    journal = {Nucl. Phys. B},
    volume = {619},
    pages = {313--358},
    year = {2001}
}

@article{Catani:1989ne,
    author = {Catani, S. and Trentadue, L.},
    title = {Resummation of the QCD perturbative series for hard processes},
    reportNumber = {DFF-93/3/89},
    journal = {Nucl. Phys. B},
    volume = {327},
    pages = {323--352},
    year = {1989}
}

@article{Yennie:1961ad,
    author = {Yennie, D. R. and Frautschi, Steven C. and Suura, H.},
    title = {The infrared divergence phenomena and high-energy processes},
    journal = {Annals Phys.},
    volume = {13},
    pages = {379--452},
    year = {1961}
}

@article{Krauss:2022ajk,
    author = {Krauss, Frank and Price, Alan and Schonherr, Marek},
    title = {YFS resummation for future lepton-lepton colliders in SHERPA},
    journal = {SciPost Phys.},
    volume = {13},
    number = {2},
    pages = {026},
    year = {2022}
}

@article{JADACH2000260,
title = {The precision Monte Carlo event generator KK for two-fermion final states in $e^+e^-$ collisions},
journal = {Comput. Phys. Commun.},
year = {2000},
volume = {130},
number = {3},
pages = {260-325},
author = {S. Jadach and B.F.L. Ward and Z. Was},
}

@article{Jadach:1991by,
    author = {Jadach, S. and Richter-Was, E. and Ward, B. F. L. and Was, Z.},
    title = {Monte Carlo program BHLUMI-2.01 for Bhabha scattering at low angles with Yennie-Frautschi-Suura exponentiation},
    journal = {Comput. Phys. Commun.},
    volume = {70},
    pages = {305--344},
    year = {1992}
}

@article{Placzek:2026hvx,
    author = {P{\l}aczek, Wies{\l}aw and Skrzypek, Maciej and Ward, Bennie F. L. and Yost, Scott A.},
    title = {\textls[+24]{Bhabha scattering at future colliders with BHLUMI/BHWIDE}},
    volume = {arXiv:2601.15265},
    journal = {arXiv},
    year = {2026}
}

@article{Hamberg:1990np,
    author = {Hamberg, R. and van Neerven, W. L. and Matsuura, T.},
    title = {A complete calculation of the order $\alpha-s^{2}$ correction to the Drell-Yan $K$ factor},
    journal = {Nucl. Phys. B},
    volume = {359},
    pages = {343--405},
    year = {1991},
}

@article{Kuraev:1985hb,
    author = {Kuraev, E. A. and Fadin, Victor S.},
    title = {On radiative corrections to $e^+e^-$ single photon annihilation at high-energy},
    journal = {Sov. J. Nucl. Phys.},
    volume = {41},
    pages = {466--472},
    year = {1985}
}

@article{Gribov:1972rt,
    author = {Gribov, V. N. and Lipatov, L. N.},
    title = {$e^+e^-$ pair annihilation and deep inelastic ep scattering in perturbation theory},
    journal = {Sov. J. Nucl. Phys.},
    volume = {15},
    pages = {675--684},
    year = {1972}
}

@article{Dokshitzer:1977sg,
    author = {Dokshitzer, Yuri L.},
    title = {\textls[-18]{Calculation of the structure functions for deep inelastic scattering and $e^+e^-$ annihilation by perturbation theory in quantum chromodynamics.}},
    journal = {Sov. Phys. JETP},
    volume = {46},
    pages = {641--653},
    year = {1977}
}

@article{Altarelli:1977zs,
    author = {Altarelli, Guido and Parisi, G.},
    title = {{Asymptotic freedom in parton language}},
    journal = {Nucl. Phys. B},
    volume = {126},
    pages = {298--318},
    year = {1977}
}

@article{Gribov:1972ri,
    author = {Gribov, V. N. and Lipatov, L. N.},
    title = {{Deep inelastic ep scattering in perturbation theory}},
    journal = {Sov. J. Nucl. Phys.},
    volume = {15},
    pages = {438--450},
    year = {1972}
}

@article{Berends:1987ab,
    author = {Berends, Frits A. and Van Neerven, W. L. and Burgers, G. J. H.},
    title = {Higher order radiative corrections at LEP energies},
    journal = {Nucl. Phys. B},
    volume = {297},
    pages = {429--478},
    year = {1988},
}

@article{FCC:2018evy,
    author = "Abada, A. and others",
    collaboration = "FCC",
    title = "{FCC-ee: The Lepton Collider}: {Future Circular Collider Conceptual Design Report Volume 2}",
    reportNumber = "CERN-ACC-2018-0057",
    journal = "Eur. Phys. J. ST",
    volume = "228",
    number = "2",
    pages = "261--623",
    year = "2019"
}

@article{FCC:2018byv,
    author = {{The FCC Collaboration}},
    collaboration = {FCC},
    title = {FCC physics opportunities: future circular collider conceptual design report volume 1},
    journal = {Eur. Phys. J. C},
    volume = {79},
    number = {6},
    pages = {474},
    year = {2019}
}

@article{CEPCStudyGroup:2018ghi,
    author = {{The CEPC Study Group}},
    collaboration = {CEPC Study Group},
    title = {CEPC conceptual design report: volume 2---physics \& detector},
    journal={arXiv},
    year={2018},
    volume={arXiv:1811.10545}
}

@article{Akhundov:2013ons,
    author = {Akhundov, Arif and Arbuzov, Andrej and Riemann, Sabine and Riemann, Tord},
    title = {The ZFITTER project},
    journal = {Phys. Part. Nucl.},
    volume = {45},
    number = {3},
    pages = {529--549},
    year = {2014}
}

@article{Blumlein:2019srk,
    author = {Blumlein, J. and De Freitas, A. and Raab, C. G. and Schonwald, K.},
    title = {The $O(\alpha^2)$ initial state QED corrections to $e^+e^-$ annihilation to a neutral vector boson revisited},
    journal = {Phys. Lett. B},
    volume = {791},
    pages = {206--209},
    year = {2019}
}

@article{Ablinger:2020qvo,
    author = {Ablinger, J. and Blumlein, J. and De Freitas, A. and Schonwald, K.},
    title = {Subleading logarithmic QED initial state corrections to $e^+e^- \rightarrow \gamma^*/{Z_{0}^*}$ to $O(\alpha^6 L^5)$},
    journal = {Nucl. Phys. B},
    volume = {955},
    pages = {115045},
    year = {2020}
}

@article{Nicrosini:1986sm,
    author = {Nicrosini, O. and Trentadue, Luca},
    title = {{Soft photons and second order radiative corrections to $e^+e^-$ ---\ensuremath{>} Z0}},
    journal = {Phys. Lett. B},
    volume = {196},
    pages = {551},
    year = {1987}
}

@article{Skrzypek:1992vk,
    author = {Skrzypek, Maciej},
    title = {{Leading logarithmic calculations of QED corrections at LEP}},
    journal = {Acta Phys. Polon. B},
    volume = {23},
    pages = {135--172},
    year = {1992}
}

@article{Cacciari:1992pz,
    author = {Cacciari, M. and Deandrea, A. and Montagna, G. and Nicrosini, O.},
    title = {{QED structure functions: a systematic approach}},
    journal = {EPL},
    volume = {17},
    pages = {123--128},
    year = {1992}
}

@article{Blumlein:2002fy,
    author = {Blumlein, Johannes and Kawamura, Hiroyuki},
    title = {$O(\alpha^2 L)$ radiative corrections to deep inelastic ep scattering},
    journal = {Phys. Lett. B},
    volume = {553},
    pages = {242--250},
    year = {2003}
}

@article{WorkingGrouponRadiativeCorrections:2010bjp,
    author = "Actis, S. and others",
    collaboration = "Working Group on Radiative Corrections, Monte Carlo Generators for Low Energies",
    title = "{Quest for precision in hadronic cross sections at low energy: Monte Carlo tools vs. experimental data}",
    eprint = "0912.0749",
    archivePrefix = "arXiv",
    primaryClass = "hep-ph",
    reportNumber = "BIHEP-TH-2009-005, BU-HEPP-09-08, CERN-PH-TH-2009-201, FNT-T-2009-03, FREIBURG-PHENO-09-07, HEPTOOLS-09-018, IEKP-KA-2009-33, LNF-09-14-P, LPSC-09-157, LPT-ORSAY-09-95, LTH-851, MZ-TH-09-38, PITHA-09-14",
    journal = "Eur. Phys. J. C",
    volume = "66",
    pages = "585--686",
    year = "2010"
}

@article{Tomalak:2022xup,
    author = {Tomalak, Oleksandr and Chen, Qing and Hill, Richard J. and McFarland, Kevin S. and Wret, Clarence},
    title = {{Theory of QED radiative corrections to neutrino scattering at accelerator energies}},
    journal = {Phys. Rev. D},
    volume = {106},
    number = {9},
    pages = {093006},
    year = {2022}
}

@article{Tomalak:2021hec,
    author = {Tomalak, Oleksandr and Chen, Qing and Hill, Richard J. and McFarland, Kevin S.},
    title = {{QED radiative corrections for accelerator neutrinos}},
    journal = {Nature Commun.},
    volume = {13},
    number = {1},
    pages = {5286},
    year = {2022}
}

@article{Hill:2016gdf,
    author = {Hill, Richard J.},
    title = {\textls[-18]{Effective field theory for large logarithms in radiative corrections to electron proton scattering}},
    journal = {Phys. Rev. D},
    volume = {95},
    number = {1},
    pages = {013001},
    year = {2017}
}

\end{document}